\documentclass[manuscript,showpacs,amsmath,amssymb,nofootinbib]{revtex4-1}
\usepackage{amsmath}
\usepackage{amssymb}
\usepackage{graphicx}
\usepackage{dcolumn}
\usepackage{bm}

\usepackage{color}

\newcommand{\notizz}[1]{{\color{black} #1}}

\begin{document}
\title{Adiabatic elimination of inertia of the stochastic microswimmer
  driven by $\alpha-$stable noise }

\author{J. Noetel$^1$, I. M. Sokolov$^1$, L.
  Schimansky-Geier$^{1,2}$\footnote{alsg@physik.hu-berlin.de}}
\affiliation{$^1$Institute of Physics, Humboldt University at Berlin,
  Newtonstr. 15, D-12489 Berlin, Germany\\$^2$Department of Physics
  and Astronomy, Ohio University, Athens, Ohio 45701, USA }

\begin{abstract}
  We consider a microswimmer that moves in two dimensions at a
  constant speed and changes the direction of its motion due to a
  torque consisting of a constant and a fluctuating component. The
  latter will be modeled by a symmetric L{\'e}vy-stable
  ($\alpha$-stable) noise. The
  purpose is to develop a kinetic approach to eliminate the angular
  component of the dynamics in order to find a coarse grained
  description in the coordinate space. \notizz{ By defining
  the joint probability density function of
  the position and of the orientation of the particle through the Fokker-Planck equation, we derive
  transport equations for the position-dependent marginal density, the
  particle's mean velocity and the velocity's variance.} At time
  scales larger than the relaxation time of the torque
  $\tau_{\phi}$ the two higher moments follow the marginal
  density, and can be adiabatically eliminated. As a result, a closed
  equation for the marginal density follows. \notizz{This equation which gives a coarse-grained description of the
    microswimmer's positions at time scales $t\gg \tau_{\phi}$, is a
  diffusion equation with a constant diffusion coefficient depending on the properties of the noise.
Hence, the long time dynamics of a microswimmer can be described as a
  normal, diffusive, Brownian motion with Gaussian
    increments.}
\end{abstract}
\pacs{05.40.-a,87.16.Uv,87.18.Tt}
\maketitle

\section{Introduction}
A popular class of models used to describe active particles assumes 
that the particles' motion remains at a constant speed. In these models the
Newtonian equations of motion for an active particle reduce to the
consideration of its orientational dynamics.
 
\notizz{The direction of motion changes due to a torque,which has a
  constant as well as fluctuating contributions, modeled as noise. The
  noise may appear due to external forces acting on the particle, due
  to interaction with other particles, or by consequences of the
  nature of the particle's internal propulsive mechanism
  \cite{romanczuk2011,gromann2016,mijalkov2013,volpe2014,geiseler2016,geiseler2016kramers,patch2017,debnath,btenhagen,babel}. The
  introduction of the constant torque is necessary to be able to
  describe situation like the ones found in some bacteria as
  Escherichia coli \cite{DiLuzio,Hill}, and in
  spermatozoa\cite{Wolley,Riedel,teeffelen,friedrich} which are known
  to swim in circles.  Artificial Janus particles
  \cite{Dhar,Kudrolli,kummel} and particles in drift chambers
  \cite{astro} can also move in circles when the symmetry is broken.
}

Active particles without noise are sometimes called microswimmers.  If
random torques are present, as in our considered cased, we call them
stochastic microswimmers.
  \notizz{Previous works on stochastic
  microswimmers have investigated the effect of random torques modeled
  by a Gaussian noise \cite{weber_sokolov_lsg}, by a dichotomous
  Markov process \cite{Weber}, and also by $\alpha$-stable
  \cite{noetl_sokolov_lsg:2017} noise. These investigations have
  focused on calculating the mean squared displacements (MSD) in dependence on the
  parameters characterizing the noise. As a result it was shown that for all
  these different kinds of angular noise the particles exhibit
ballistic motion at short time scales and diffusive motion at longer
times.  The ballistic motion is caused by the inertia of the swimmer,
i.e. due to the fact that the particle remembers for a certain time
interval the direction it has currently moved in. However, for longer
times, i.e. at time scales larger than the relaxation time of the
orientation, this orientational memory fades out, and the normal
diffusive behavior sets on.
}

Such a crossover between ballistic and diffusive motion is best known
for normal (i.e. non-active) Brownian motion
\cite{langevin,Becker52,Hwalisz,Milster}. The particle's motion is
described by the joint probability density in the phase space, i.e.
for the particle's velocity and position. The coarse-grained
description for the position variables only leads to a diffusion
equation.  Therein, the velocity of the particle as a dynamical
variable has been eliminated. \notizz{This coarse-grained description 
is valid at time scales
  $\Delta t\gg \tau_{v}$ and at length scales $|\Delta \vec{r}| \gg
  l_{v}=\tau_{v} \sqrt{k_{\rm B}T/m}$.  Therein $\tau_{v}=m/\gamma$ is
  the velocity relaxation time and $l_{v}$ is the brake path, or
  persistence length of Brownian motion, with $m$ being the mass of the
  particle and $\gamma$ the particle's friction coefficient.  
  In situations where $\tau_v$ is small the dynamics is often
  referred to as an overdamped dynamics.

  For Brownian motion, there exists a vast literature which considers
  the elimination of inertia at larger time and length scales.
  Already Kramers in his seminal work \cite{Kramers} found an elegant
  way to eliminate the velocity. Using the factorizing properties of
  the Fokker-Planck operator, he was able to derive the diffusion
  equation for the marginal probability density $\rho(\mathbf{r},t)$
  (see also \cite{Becker52}).  Later on, many other approaches have
  been formulated, including the projection operator formalism and
  approaches which adiabatically eliminate variables
  \cite{haken,gardiner,gardiner:82,ucna,lsg_talkner,sancho}.  }

Here, we seek for the foundation of the diffusive motion of the
microswimmer by adiabatic elimination of the angular inertia.
\notizz{We will derive a coarse-grained description of the particle's
  motion at time and length scales where the angular memory fades out,
  and calculate the distribution of its displacements. To the best of
  our knowledge, such an elimination procedure for a stochastic
  microswimmer has not been previously considered in detail. The case
  in which Gaussian white noise models the torque was
  elaborated only recently. Therein, the angular inertia was
  eliminated in the corresponding Langevin equation, and the coarse
  grained Langevin equations of the diffusing microswimmer were
  derived and discussed \cite{Milster}. For sake of completeness, we
  show in Appendix \ref{append1} the elimination with torque and
  Gaussian white noise, from the Langevin equations.}

\notizz{In the present work we will study the broader situation, and
  consider a model for a stochastic microswimmer in which
  orientational changes are due to a combination of a constant torque
  and of random fluctuations described by an $\alpha$-stable noise.}
We are interested in obtaining the \notizz{coarse grained} dynamical
description in the coordinate space.  Surprisingly, despite the
L{\'e}vy nature of the noise in the angular variable, the coarse
grained dynamics is found to be modeled through Gaussian white noise
acting in the coordinate space. Thus, the long time behavior can be
universally described as Brownian motion.

\notizz{We have been unable to perform this elimination on the level
  of the Langevin equation. Due to the heavy-tailed nature of the
  increments of the $\alpha$-stable noise, the moments needed in the
  corresponding derivation do not exist. Here we will use a kinetic
  approach which, for the case of Gaussian white noise, has been
  developed in \cite{Milster}.  \notizz{This approach uses the
    transport equations for the first three moments of the velocity
    components.} These velocity components are expressed by the cosine
  and sine of the orientation.  Since these are bounded functions,
  their mean values exist even for the case that the angular dynamics
  is due to {L\'evy} noise.}

In section \ref{sec_tor_white} we introduce the model and present the
results of simulations of the stochastic microswimmer. In section
\ref{sec_gauss} we formulate the Fokker-Planck equation for the joint
probability density function (pdf) of the orientation and the position
of the microswimmer in two dimensional coordinate space. We then
derive equations for the reduced moments of this pdf, which are the
marginal pdf of the position of the particle, the average velocity
components, and their variances. By expressing higher moments through
the first three, these equations represent an approximation of the
dynamics in the position space. In section \ref{sec_adi_eli} we
discuss the procedure of the adiabatic elimination of the velocity
variables. At time scales larger than the crossover time $\tau_{\phi}$, we
assume that the variance of the velocity follows the marginal density
and the velocities squared. Further on, we assume that the mean
velocities follow the marginal density, and eliminate both the
velocity's first and second moment.  As a result we obtain a closed
equation for the marginal position-dependent density. This is a
diffusion equation \notizz{having as solution the Gaussian
  distributions of independent spatial displacements. The diffusion
  coefficient characterizing the linear growth of the variance of
  these displacements} is a constant that depends on the parameters
characterizing the $\alpha$-stable noise.

\section{The Model}
\label{sec_tor_white}
We consider a microswimmer in two dimensions whose position is given
by a vector $\vec{r}=\left(x(t),y(t)\right)$. The position space is
unbounded.  The swimmer starts at time $t_0=0$ at position
$\vec{r}_0=(x_0=0,y_0=0)$ and has a constant speed $v_0>0$ but the
actual heading is given by the angle $\phi(t) \in [0,2\pi) $, so that
the dynamics is described by a set of equations
\begin{eqnarray} 
\frac{\text{d}\vec{r}}{\text{d}t}=\vec{v}=v_0
\begin{pmatrix}
         \cos\phi(t) \\ \sin\phi(t)
        \end{pmatrix}\,,
\label{r_dot}
\end{eqnarray}
\begin{eqnarray} 
\frac{\text{d}\phi}{\text{d}t}=\Omega+\frac{\sigma}{v_0}\xi(t)\,.
\label{phi_dot}
\end{eqnarray}
Here $\Omega$ is a time-independent torque, and the noise $\xi(t)$ is
assumed to be white, $\alpha$-stable and symmetric \cite{feller}, 
with noise strength $\sigma$.

The term $\alpha-$stable (L\'evy stable) refers to the fact that the
sum of random variables following a stable distribution follows the
same probability distribution up to a rescaling and a re-centering.
The increments of the noise are independent.
The commonly used Gaussian noise (corresponding to $\alpha=2$) is one
example of such stable white
noises. 
The parameter $\alpha$ is the L\'evy stability index, and
characterizes the tails of the probability distribution.  Decreasing
$\alpha$ leads to larger sudden changes in the direction of motion.
For values of $\alpha<2$ the tails of the corresponding probability 
distributions become heavier, and the corresponding random variables show more outliers. 
Distributions with $\alpha<2$ have infinite variance and for
$\alpha\leq1$ even the first moment no longer exist.

Such non-Gaussian noise distributions are quite useful to describe the
motion of entities which behave like hunted rabbits or
antelopes. Typical trajectories of such a running animal with rapid
turning behavior are not best described by Gaussian
increments. \notizz{ The motion of a particle under such conditions is
  better characterized by a run and tumble behavior
  \cite{runandtumble1,runandtumble2} with fast periods of tumbling.
  Recent experiments on motions of fruit flies
  \cite{Kim2017,Jundi2017} have reported similar trajectories with
  rapid changes.}

The active particles described by the set of equations (\ref{r_dot}),
(\ref{phi_dot}) perform a ballistic motion on a short time-scale
$t\ll\tau_\phi$ \cite{noetl_sokolov_lsg:2017} with
\begin{equation}
\tau_\phi=\left(\frac{v_0}{\sigma}\right)^\alpha,
\label{eq:relaxt}
\end{equation} 
being the relaxation time and with the persistence length
$l_p=v_0\tau_\phi$. At longer times $t\gg\tau_\phi$ and at distances
$|vec{r}| \gg l_p$ the motion becomes diffusive with the effective
diffusion coefficient
\begin{equation}
D_\text{eff}=\frac{v_0^2}{2}\frac{\tau_\phi}{1+\left(\Omega\tau_\phi\right)^2}.
\label{eq:deff}
\end{equation} 
The diffusion coefficient and the relaxation time were calculated in
\cite{noetl_sokolov_lsg:2017} using the Green-Kubo relation.
\begin{figure}[ht]
  \includegraphics[width=.7\textwidth]{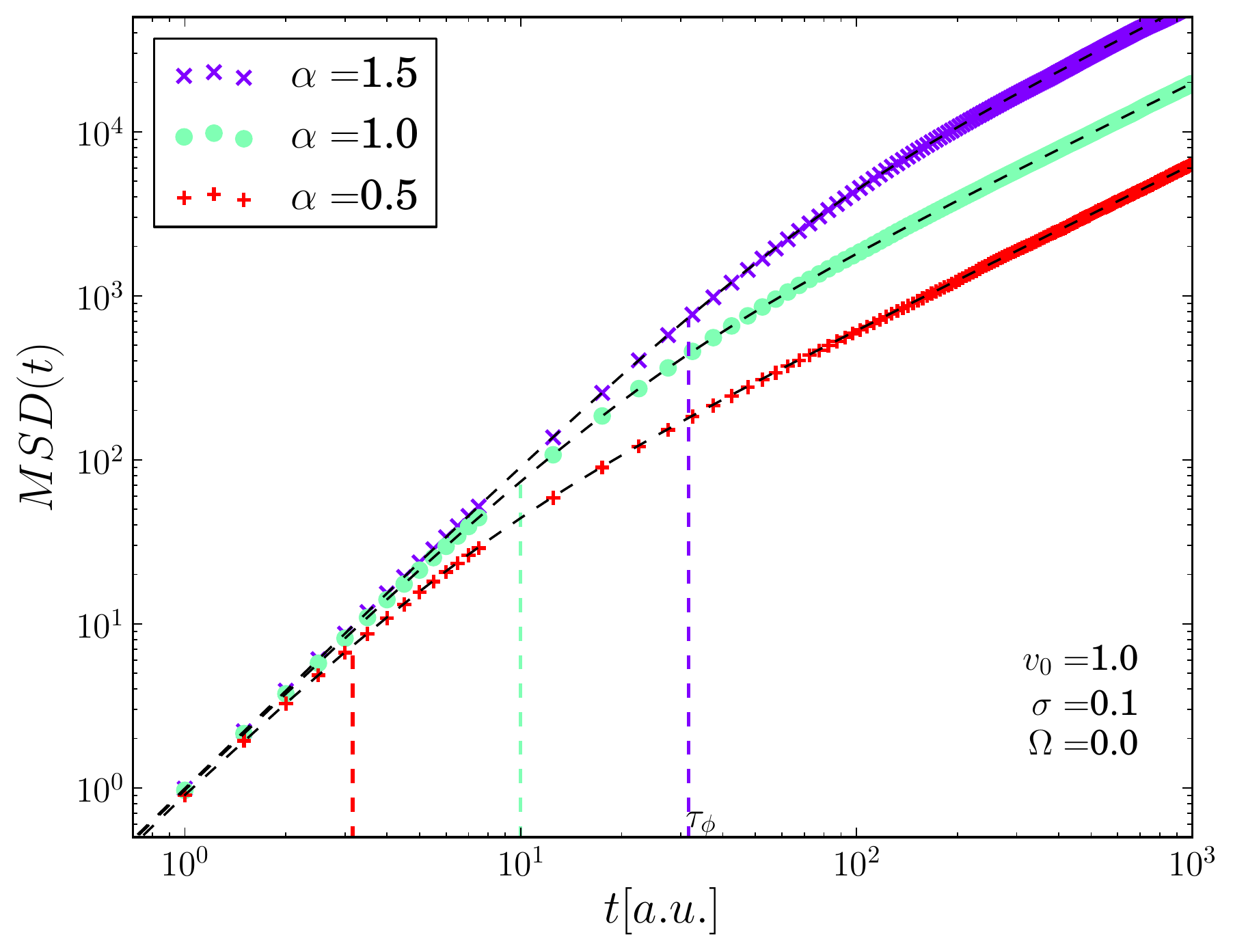}
  \caption{Mean squared displacement of the stochastic microswimmer
    for various values of $\alpha$. The crossover between ballistic and
    diffusive behavior takes places at indicated transition times. The transition times equal 
    the relaxation times of the angular dynamics $
    \tau_{\phi}$ and depend on the choice of $\alpha$. The broken line corresponds to the MSD calculated using the
    Green-Kubo relation in \cite{noetl_sokolov_lsg:2017}.}
   \label{fig:msd}
\end{figure}

\notizz{The mean squared displacement as a function of time is
  shown in Figure \ref{fig:msd} for various values of $\alpha$. Due to
  the angular inertia, or memory, the motion is ballistic for times
  smaller $t\ll\tau_\phi$ and at times $t \gg \tau_\phi$ becomes
  diffusive with the diffusion coefficient $D_\text{eff}$
  given by eq.\eqref{eq:deff}. In the figures the crossover time, being the relaxation
  time of the angular dynamics, increases with $\alpha$ when other parameters are kept fixed. }
\begin{figure}[h]
  \includegraphics[width=.45\textwidth]{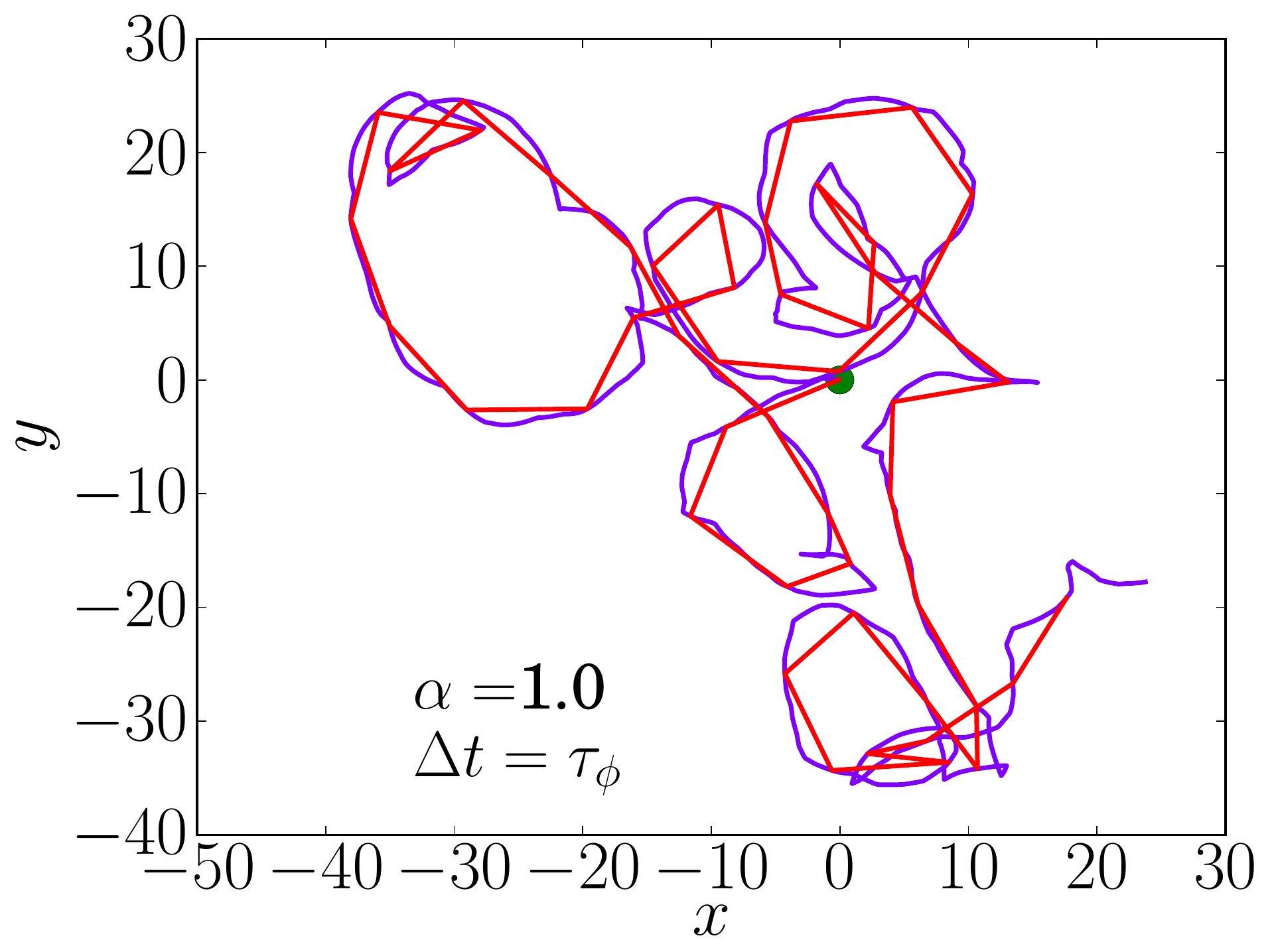}
  \includegraphics[width=.45\textwidth]{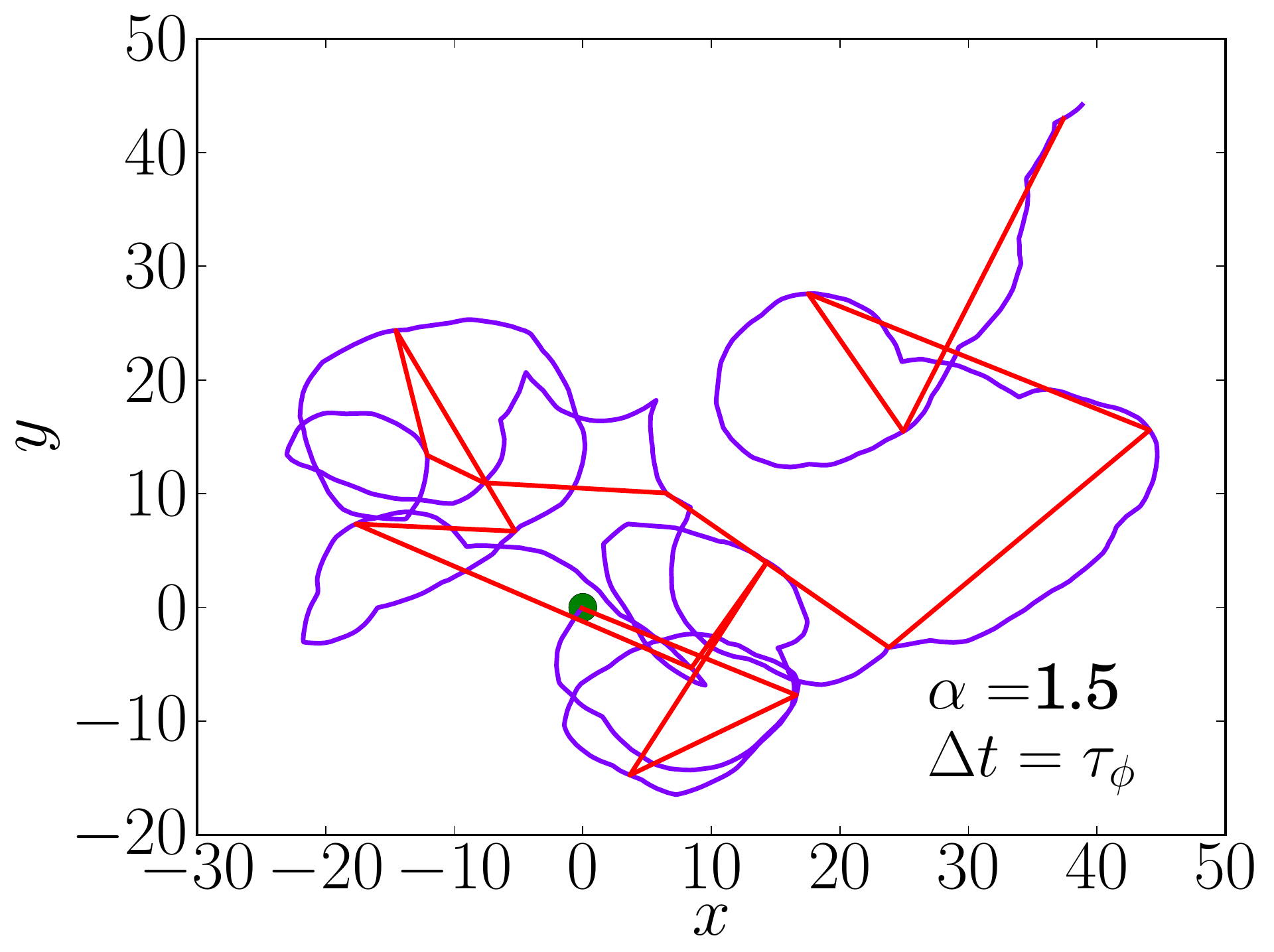}
  \caption{\notizz{Sample trajectories of the stochastic microswimmer
      with constant speed for $\alpha=1$ (left) and $\alpha=1.5$
      (right). Trajectories start at $(0,0)$, marked as green dot. The
      blue trajectories are the persistent paths. Their coarse grained
      dynamics are shown in red and
      taken in sampling intervals of $\Delta t= \tau_{\phi}$. Other
      parameters $v_0= 1.0$, $\sigma=0.1$ and $\Omega=0.1$. The trajectories contain a total simulation time of t=500.}}
   \label{fig:traj1}
\end{figure}

\notizz{Figure \ref{fig:traj1} shows sample trajectories of stochastic
  microswimmers for $\alpha=1$ (left panel) and for $\alpha=1.5$
  (right panel). The blue lines show simulated paths plotted with a
  time increment $\Delta t=0.1$ much smaller than the relaxation time.
  One can see motion over almost straight lines interrupted by sharp
  turns.  For lower values of $\alpha$, here $\alpha=1$, trajectories
  become rather smooth occasionally interrupted by sharp turns. In
  contrast, the red paths represent the coarse grained dynamics of
  the blue trajectories sampled with $\Delta t= \tau_{\phi}$ (compare
  Eq.\eqref{eq:relaxt}). At this large time scale the motion is just at the crossover time, 
  beyond which the motion starts to be diffusive with
  independent increments and statistically indistinguishable from
  Brownian motion.}

\notizz{For all values of $\alpha$}, an initial angle
$\phi(t_0=0)=\phi_0$ is forgotten after the relaxation time $\tau_\phi$
has elapsed. Moreover, for long enough times $t\gg\tau_\phi$ the
angular transition probability density $Q=Q(\phi,t|\phi_0,t_0)$ becomes
uniform.  This follows from the corresponding Fokker-Planck-Equation
(FPE) for $Q$ which decouples from the coordinate dynamics since
Eq.(\ref{phi_dot}) is autonomous. This FPE reads
\cite{Ditlevsen,Schertzer}
\begin{eqnarray} 
\frac{\partial}{\partial t}\,Q(\phi,t|\phi_0,t_0)
\,=\,-\,\frac{\partial}{\partial \phi}\,\Omega
Q\,+\,\frac{1}{\tau_\phi}\frac{\partial^\alpha}{\partial
  |\phi|^\alpha}\, Q\,,
\label{fpe_angle}
\end{eqnarray}
with the $\alpha$-th symmetric Riesz-Weyl fractional derivative
\begin{eqnarray} 
  \frac{\partial^\alpha}{\partial|\phi|^\alpha}\,Q(\phi,t|\phi_0,t_0)\,
  =\,-\,\frac{1}{2\pi}\,\int_{-\infty}^{\infty}\,{\rm
    d}k\,e^{-ik\phi}\,|k|^\alpha \,Q(k,t|\phi_0,t_0)\,,
\end{eqnarray}
where 
\begin{eqnarray} 
  Q(k,t|\phi_0,t_0)\, =\,\int_{-\infty}^{\infty}\,{\rm d}\phi e^{ik\phi}\,Q(\phi,t|\phi_0,t_0)\,,
\end{eqnarray}
is the Fourier transform of the transition probability density with
respect to the angular variable $\phi$.  The solution to eq. \eqref{fpe_angle} is
\begin{eqnarray} 
Q(\phi,t|\phi_0,t_0=0)=\frac{1}{\pi}\left(\frac{1}{2}+\sum_{n=1}^\infty\cos\left(
n(\phi-\phi_0-\Omega t) \right)e^{-\frac{ n^\alpha
    t}{\tau_\phi}}\right),
\label{eq:pphi}
\end{eqnarray}
which takes into account the $2\pi$-periodicity of the angular
variable and the initial condition $\phi(t=0)=\phi_0$, i.e.
$Q(\phi,t=0|\phi_0,0)=\delta(\phi-\phi_0)$. The case $\alpha=2$
corresponds to a Gaussian white noise. For times $t\gg\tau_\phi$ the
probability density to find a specific angle becomes constant
$1/(2\pi)$ for all symmetric $\alpha-$stable noise types.
\notizz{Note that for the slowest mode the only parameter which depends on the noise type 
is the relaxation time $\tau_\phi$.  The flattening of the angular
  distribution is the key feature that allows the description of our
  active particles as Brownian particles in the long-time limit.}

As already reported in \cite{noetl_sokolov_lsg:2017}, numerical
simulations show a Gaussian distribution of displacements at times
larger the relaxation time $t\gg\tau_\phi$ for all considered noise
types. In polar coordinates $(r,\varphi)$ the displacement's
distribution is independent from the spatial angle $\varphi$ at this
time scale. For simplicity, we take the origin as the initial
position, i.e.  $x_0=0,y_0=0$. The distribution of the distance to
the origin $r(t)=\sqrt{x^2(t)+y^2(t)}$ is given by the Rayleigh
distribution.  This is to be expected since for longer time increments
of the heading become uncorrelated, and the step
length is finite.
\begin{figure}[ht]
  \includegraphics[width=.7\textwidth]{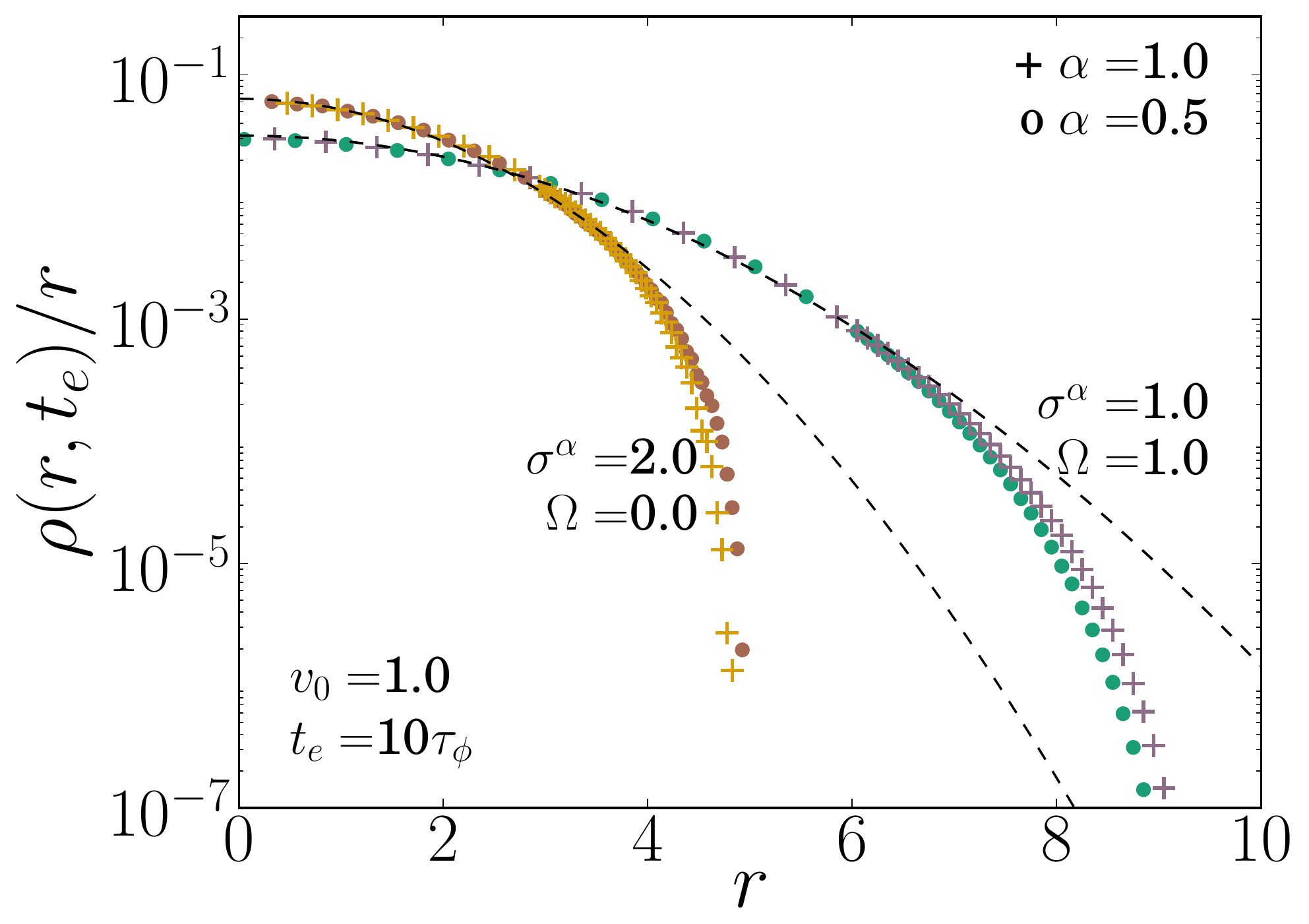}
  \caption{\notizz{Simulation results for the radial displacement
      distribution divided by $r$, i.e. $\rho(r,t_e|0,0)/r$ taken at
      time $t_e =10 \tau_\phi$. Results are shown for two different
      values of $\alpha$ indicated by circles ($\alpha=0.5$) and
      crosses ($\alpha=1.0$). Results are for two different noise intensities
      with indicated values. Note, due to the dependence of $\tau_\phi$
      on $\alpha$ the brown and green results have been taken at
      different end time $t_e$. Black dashed line corresponds to a
      Rayleigh distribution with mean squared displacement $\langle
      r^2 \rangle=4D_{\text{eff}}t$ and $D_{\text{eff}}$ from
      \eqref{eq:deff}.  Other parameters are given in the figure.} }
   \label{fig:traj}
\end{figure}

Figure \ref{fig:traj} shows simulation results for the radial
displacement distribution divided by $r$, i.e. for
$\rho(r,t_e|0,0)/r$, after time $t_e=10\tau_\phi$. Black dashed line
corresponds to a Rayleigh distribution with mean squared displacement
$\langle \, r^2 \,\rangle=4D_{\text{eff}}t$ where $D_{\text{eff}}$ is
Eq.~\eqref{eq:deff}. The Rayleigh distribution coincides well with
the simulations up to $r \approx 0.7 v_0 t_e$. When $r >> 0.7 v_0 t_e$ the inertia and
the constant kinetic energy of the microswimmer cause deviations.

\section{Kinetic approach of eliminating the angular memory}
\label{sec_gauss}
In this section we will follow a kinetic approach based on the
Fokker-Planck equation, similar to the approach used in \cite{Milster}
for active particles with Gaussian white noise. Here we consider the
broader class of symmetric $\alpha-$stable noises. In addition, we
consider the effects of the constant torque, a physical situation
often met for motile particles.

The kinetic approach is based on the first three reduced
coordinate-dependent moments of the velocity, which we will soon
define. In our two dimensional situation with constant speed the
reduced moments are obtained by averaging over the heading $\phi$.
Afterwards as in kinetic theory, the spatio-temporal evolution is
determined by the transport equations for the marginal probability
density, components of the the average velocities and the variances of
the velocity.  In these equations, we will adiabatically eliminate
higher moments. This enables us to approximate the variances and mean
velocity in order to find a closed description for the dynamics of
active particle considered.

Associated with the given Langevin equations
\eqref{r_dot},\eqref{phi_dot} is the following FPE \cite{Ditlevsen,
  Schertzer}:
\begin{eqnarray} 
\frac{\partial}{\partial
  t}P(x,y,\phi,t|x_0,y_0,\phi_0,t_0)=-\vec{v}\cdot\nabla
P-\frac{\partial}{\partial \phi}\Omega
P+\frac{1}{\tau_\phi}\frac{\partial^\alpha}{\partial |\phi|^\alpha}P,
\label{fpe}
\end{eqnarray}
$\vec{v}=(v_x,v_y)^T$ for the transition pdf $P=P(x,y,\phi,t|x_0,y_0,\phi_0,t_0)$.
The aforementioned reduced moments are the marginal probability
density $\rho(x,y,t)$, the mean velocity components $u_i(x,y,t)$ and
the variances $\sigma_{ij}(x,y,t)$
\begin{equation}
\rho(x,y,t)=\int_{0}^{2\pi} P(x,y,\phi,t)d\phi
\label{eqrho}
\end{equation}
\begin{equation}
\rho(x,y,t)u_i(x,y,t)=\int_{0}^{2\pi} v_iP(x,y,\phi,t)d\phi
\label{equ}
\end{equation}
\begin{equation}
\rho(x,y,t)\sigma_{ij}^2(x,y,t)=\int_{0}^{2\pi} (v_i-u_i(x,y,t))(v_j-u_j(x,y,t))P(x,y,\phi,t)d\phi
\label{eqsigma}
\end{equation}
with $i,j \in [x,y]$. We omitted for readability the conditions of $P$
in the equations \footnote{Since the pdf in Eq.(\ref{fpe}) is the
  transition probability density function and conditioned to the
  initial state, the reduced moments are conditioned moments as well.
  In kinetic theory this condition is reflected by the formulation of
  corresponding initial conditions for the three moments.}.

\notizz{The application of this kinetic approach to our model is
  straightforward since such properties as the non-existence of higher
  moments $\langle \phi^n \rangle$ (for $n\ge 2$ with $\alpha<2$ and
  for $n\ge 1$ in case of $\alpha\le 1$) do not matter.  While these
  moments might not exist, the mathematical expectations of the periodic
  functions $\langle \cos(\phi) \rangle$,$\langle \sin(\phi) \rangle$
  and their powers stay finite, see Appendix \ref{frac_mom} for details.}

The transport equations for the corresponding moments are derived by
differentiating equations \eqref{eqrho},\eqref{equ},\eqref{eqsigma}
with respect to time and using the Fokker-Planck equation \eqref{fpe}
which has to be integrated over the angle $\phi$. As a result, we obtain
the continuity equation
\begin{equation}
\frac{\partial}{\partial t}\rho\,=\,-\frac{\partial}{\partial x} \rho
u_x \,-\,\frac{\partial}{\partial y} \rho u_y\,,
\label{eq:rhot}
\end{equation}
and the transport equation for the momentum
\begin{equation}
  \frac{\partial}{\partial t}\rho u_i=
  -\left(\frac{\partial}{\partial x_i}
  \rho(u_i^2+\sigma_{ii}^2) +\frac{\partial}{\partial x_j}\rho(u_i
  u_j+\sigma_{ij}^2) \right)-a_i\Omega \rho u_j -\frac{1}{\tau_\phi}
  \rho u_i\,.
\label{eq:ut}
\end{equation}
Differentiating \eqref{eqsigma} with respect to time, we obtain the
balance equations for the variances
\begin{equation}
\frac{\partial}{\partial
  t} \rho \left(u_i^2+\sigma_{ii}^2\right)=A_{ii}-\frac{2^\alpha}{\tau_\phi}\rho \left(u_i^2+\sigma_{ii}^2-\frac{v_0^2}{2}
\right)-a_i2\Omega\rho \left(u_iu_j+\sigma_{ij}^2\right)\,,
\label{eq:sigmaii}
\end{equation}
and, respectively, for the covariance; 
\begin{equation}
\frac{\partial}{\partial t}\rho \left(u_iu_j+\sigma_{ij}^2\right)=A_{ij}-\frac{2^\alpha}{\tau_\phi}\rho \left(u_iu_j+\sigma_{ij}^2\right)+\Omega \rho\left(a_i\left(u_i^2+\sigma_{ii}^2 \right)+a_j \left(u_j^2+\sigma_{jj}^2\right)\right)\,.
\label{eq:sigmaij}
\end{equation}
Notably, the specific noise distributions of the $\alpha$-stable noise
sources enter all transport equations only through constant
parameters, i.e. through the value of $2^\alpha$ and through
$\tau_\phi$.  In the equations above \eqref{eq:ut},\eqref{eq:sigmaii}
and \eqref{eq:sigmaij} the indices run over $i,j=x,y $ and $i\ne j$.
Also, we do not sum over a repeated indices. Further on, $a_x=1$, 
$a_y=-1$ and the appearing higher moments $A_{ij}$ take the
following form:
\begin{equation}
\begin{split}
A_{xx}=-\frac{\partial}{\partial x}\int_{0}^{2\pi}v_0^3\cos^3(\phi)P(x,y,\phi,t)d\phi-\frac{\partial}{\partial y}\int_{0}^{2\pi}v_0^3\cos^2(\phi)\sin(\phi)P(x,y,\phi,t)d\phi 
\end{split}
\end{equation}
\begin{equation}
\begin{split}
A_{yy}=-\frac{\partial}{\partial x}\int_{0}^{2\pi}v_0^3\cos(\phi)\sin^2(\phi)P(x,y,\phi,t)d\phi-\frac{\partial}{\partial y}\int_{0}^{2\pi}v_0^3\sin^3(\phi)P(x,y,\phi,t)d\phi 
\end{split}
\end{equation}
\begin{equation}
\begin{split}
A_{xy}=-\frac{\partial}{\partial x}\int_{0}^{2\pi}v_0^3\cos^2(\phi)\sin(\phi)P(x,y,\phi,t)d\phi-\frac{\partial}{\partial y}\int_{0}^{2\pi}v_0^3\cos(\phi)\sin^2(\phi)P(x,y,\phi,t)d\phi 
\end{split}
\end{equation}

\section{Adiabatic elimination}
\label{sec_adi_eli}
We will now adiabatically eliminate the angular inertia, or angular
memory from the transport equations \eqref{eq:sigmaii}.
\notizz{In contrast to the case of Brownian particles, the relaxation time does depend
  on the noise intensity, $\tau_{\phi} \propto 1/\sigma^\alpha$ .
  Hence, large noise intensity $\sigma$ corresponds to larger angular
  variability, and therefore to faster relaxation. We will consider
the limit of large noise meaning that the relaxation time
  $\tau_{\phi}$ can be considered as small compared to the time scale
  of observation.} Multiplication of \eqref{eq:sigmaii} and
\eqref{eq:sigmaij} by $\tau_{\phi}/2^{\alpha}$ yields an expression
that allows us to neglect the time derivatives and the higher moments
in $A_{ii}$ and $A_{ij}$ in these equations.

We
retain terms with $\Omega \tau_{\phi}$ since $\Omega$ could be large
in contrast to the derivatives of higher moments which are considered
small.

Therefore, we neglect the temporal derivatives in the equations for
variances and covariances.  Also the influence of the higher moments
disappears in the limit of small $\tau_{\phi}$. The variances as well
as the covariances then follow the other time dependent moments.
Starting from Eq.\eqref{eq:sigmaii} we derive the following two
equations:
\begin{equation}
\rho\left(u_x^2+\sigma_{xx}^2\right)\, + \,2\Omega\frac{\tau_\phi}{2^\alpha}\,\rho \left(u_xu_y+\sigma_{xy}^2\right) \,=\,\rho \frac{v_0^2}{2} \,+\,\mathcal{O}(\tau_{\phi})
\label{eq:sigmaii2x}
\end{equation}
\begin{equation}
\rho\left(u_y^2+\sigma_{yy}^2\right)\, -\,2\Omega\frac{\tau_\phi}{2^\alpha}\,\rho \left(u_xu_y+\sigma_{xy}^2\right) \,=\,\rho \frac{v_0^2}{2} \,+\,\mathcal{O}(\tau_{\phi}).
\label{eq:sigmaii2y}
\end{equation}
 Respectively,  using Eq.~\eqref{eq:sigmaij}  we obtain 
\begin{equation}
 \Omega\frac{\tau_\phi}{2^\alpha}\rho\,\left(u_y^2+\sigma_{yy}^2-u_x^2- \sigma_{xx}^2 \right)\,+\, \rho\left( u_x u_y+\sigma_{xy}^2\right)\,=\,0\,+\,\mathcal{O}(\tau_{\phi})\,.
\label{eq:sigmaij2}
\end{equation}
The solution of this set of three bi-linear equations is:
\begin{equation}
u_i u_j+\sigma_{ij}^2 \,=\,\delta_{ij}\frac{v_0^2}{2} \,+\,\mathcal{O}(\tau_{\phi})\,.
\label{eq:sigmaijf}
\end{equation}
This solution is a kind of equipartition theorem for the two terms in the kinetic
energy and is valid at times larger the relaxation time.

Next we consider the momentum balance, Eq.~\eqref{eq:ut}. Assuming again
$t\gg\tau_\phi$ will allow us again to neglect the time
derivative. Further on, inserting therein the values of the
(co-)variances as given by Eq.~\eqref{eq:sigmaijf} yields for $i,j=x,y; i\ne j$
\begin{equation}
  \rho \left( u_i\,+\, a_i\Omega \tau_\phi u_j\right) \,=
  \,\tau_{\phi} \frac{v_0^2}{2} \,\frac{\partial}{\partial x_i}
  \rho\,+\,\mathcal{O}(\tau_\phi),
\end{equation}
meaning that the two first momenta follow quickly the marginal
density. Like $\Omega$, the velocity $v_0$ is arbitrary, and therefore  $v_0^2
\tau_{\phi}$ is not negligible in general (see Eq.~\eqref{eq:deff}). The
solution with respect to the components of the mean flux can be
easily obtained. It gives
\begin{eqnarray}
&&\rho u_x\,=\, -\,\frac{\tau_{\phi}}{1+\Omega^2
    \tau_{\phi}^2}\frac{v_0^2}{2}\,\left(\frac{\partial}{\partial
    x}\rho\,-\,\Omega\tau_{\phi}\, \frac{\partial}{\partial y} \rho
  \right)\,+\,\mathcal{O}(\tau_\phi)\,,  \nonumber \\ &&\rho u_y\,=\,
  -\,\frac{\tau_{\phi}}{1+\Omega^2
    \tau_{\phi}^2}\frac{v_0^2}{2}\,\left(\Omega\tau_{\phi}\,
  \frac{\partial}{\partial x}\rho\,+\,\frac{\partial}{\partial y} \rho
  \right)\,+\,\mathcal{O}(\tau_\phi)\,.
\label{eq:fluxes}
\end{eqnarray}
Eventually, we put these expressions for the mean momenta into the
continuity equation \eqref{eq:rhot}. Hence, up to first order of
$\tau_\phi$ we derive the evolution equation for the marginal
probability density $\rho(x,y,t)$ of a stochastic microswimmer driven
by $\alpha$-stable noise. It is the well known diffusion equation
reading
\begin{equation}
\frac{\partial}{\partial t}\rho(x,y,t)\,=\,D_{\rm
  eff}\left(\frac{\partial^2}{\partial x^2}+\frac{\partial^2}{\partial
  y^2}\right)\rho\, +\,\mathcal{O}(\tau_{\phi})\,.
\label{eq:rho_f}
\end{equation}
This is the desired dynamics of the marginal probability density for the
position, or the displacement for the coarse grained  micro-swimmer up to
first order of $\tau_{\phi}$. In this regime, which
is established at times longer than the relaxation time
$\tau_\phi=(v_0/\sigma)^{\alpha}$, the specific characteristics of the
selected $\alpha$-stable noise enters the time evolution only through
the diffusion coefficients $D_{\text{eff}}$ as given by Eq.~\eqref{eq:deff}.

\section{Discussion and conclusions}
\label{sec_disc}
In this section we will discuss our findings. We start our discussion 
with the result for the variances \eqref{eq:sigmaijf}, afterwards we
discuss the continuity equation \eqref{eq:rhot}, and then
the validity of the diffusion approximation \eqref{eq:rho_f}. For
the latter we will also compare the approximation with simulations of
the initial system.

i) The position dependence of the variances $\sigma_{ij}(x,y,t)$ is
coupled to the mean velocities by \eqref{eq:sigmaijf}.

 \notizz{ If the heading directions have reached
  equilibrium $P_0(\phi)=1/(2\pi)$, the position-dependent ensemble-averaged
  velocities $u_i(x,y,t)$ practically vanish and the variances
  become $\sigma^2_{ij}(x,y,t)=\delta_{ij}v_0^2/2$. 
  
  There exists a maximal distance $v_0 t$ a particle can travel during
  the time interval $t$.  The particles that have moved over such a
  distance have hardly changed the direction of their motion: the
  averaged absolute velocities for particles close to the maximal
  distance will be almost $v_0$, and the variances vanish. Hence, the
  approximation that the variances $\sigma_{ii}^2$ are independent
  from the velocity components $u_j$, with $i\ne j$, does not hold
  anymore.

Thus, we expect the approximation for the (co-)variances \eqref{eq:sigmaijf} to be valid for
$|\vec{r}|\ll v_0 t$.}

ii) The microswimmers considered is this paper have constant speed;
the kinetic energy is constant. The conservation of energy is
expressed through the relation
\begin{equation}
\sigma_{xx}^2\rho+\sigma_{yy}^2\rho+u_x^2\rho+u_y^2\rho=v_0^2\rho\,,
\label{eq:kons}
\end{equation}
which is derived from the definition of our velocities \eqref{r_dot}
and the variances \eqref{eqsigma}.  Our results for the variances
\eqref{eq:sigmaijf} obey this conservation of energy.  The equations
for the variances express the equipartition theorem. In our two
dimensional system every degree of freedom acquires half of the
available energy $\langle E_{ii}\rangle=\sigma^2_{ii}/2$, and the
mixed component vanishes for particles far away from the maximal
distance $|\vec{r}|\ll v_0 t$.

iii) Taking the second derivative of the marginal
density with respect to time and using the continuity equation \eqref{eq:rhot} yields
\begin{equation}
  \frac{\partial^2}{\partial t^2}\rho\,=\, -\frac{\partial}{\partial x} \,\frac{\partial}{\partial t} \rho u_x\,-\,\frac{\partial}{\partial y}\frac{\partial}{\partial t}\rho u_y \,.
  \label{eq:tele1}
\end{equation}  
This second derivative stands for the effects of inertia since it is
dominated by the derivative of the momentum flux.  Now, we insert in
the r.h.s. the expression \eqref{eq:sigmaijf} which represents the
equipartition of the kinetic energy, to obtain a telegrapher's-like
equation
\begin{equation}
  \frac{\partial^2}{\partial t^2}\rho\,+\,\frac{1}{\tau_{\phi}}
  \frac{\partial}{\partial t}\rho\, =\,\frac{v_0^2}{2}\left(\frac{\partial^2}{\partial x^2}\,+\,\frac{\partial^2}{\partial y^2}\right) \rho -\Omega \left( \frac{\partial}{\partial x} \rho u_y - \frac{\partial}{\partial y} \rho u_x\right) \,.
  \label{eq:tele2}
\end{equation}  
Using Eqs. \eqref{eq:fluxes} we can combine the two terms on the
r.h.s. Multiplying both parts of the ensuing equation by $\tau_{\phi}$
we finally obtain
\begin{equation}
\tau_{\phi}\frac{\partial^2}{\partial t^2}\rho\,+\,\frac{\partial}{\partial t}\rho\, =\,\frac{v_0^2}{2} \frac{\tau_{\phi}}{1 + \Omega^2 \tau_{\phi}^2}\left(\frac{\partial^2}{\partial x^2}\,+\,\frac{\partial^2}{\partial y^2}\right) \rho.
\label{eq:tele3}
\end{equation} 
Given Eq.~\eqref{eq:fluxes} and from the assumption of the equipartition of kinetic energy, implies $\tau_{\phi}$ to be small. 
Hence, we see
that the second temporal derivative of the density in
Eq.~\eqref{eq:tele3} is preceded by a small numerical factor and can
be neglected.  Therefore the inertia part can be dropped, yielding the
diffusion approximation Eq.~\eqref{eq:rho_f}.

iv) The diffusion approximation Eq.~\eqref{eq:rho_f} is valid for
times $t\gg\tau_{\phi}$, and for displacements $x^2+y^2\ll v_0^2t^2$:
\notizz{The first inequality is necessary for the relaxation
of the heading angles to a homogeneous distribution. The second inequality specifies in which spacial 
area the relaxation happens.
} 

The coarse grained dynamics of our system corresponds to a Brownian
motion with the diffusion coefficient given by Eq.~\eqref{eq:deff}.
The properties of the noise distribution (e.g.  its stability index
and intensity) enter only in the diffusion coefficient.  For times
large enough, $t \gg \tau_{\phi}$, when correlations are lost, it does
not matter whether a particle performs a lot of small turns or fewer
huge ones; the heading always becomes uniformly distributed in an
interval $[0,2\pi)$.  As expected from the central limit theorem, the
displacement distribution then becomes Gaussian, since the spatial
increments become independent from one another and have a maximal
length in a given time interval, i.e. do possess the finite second
moment.  Changing from Cartesian $(x,y)$ to polar coordinates
$(r,\varphi)$, the solution of the equation for the displacement
\eqref{eq:rho_f} corresponds to the Rayleigh-distribution
\begin{equation}
\rho(r,\varphi,t)=\frac{r}{4\pi D_\text{eff}t}e^{-\frac{r^2}{4D_\text{eff}t}},
\label{eq:rayl_varphi}
\end{equation}
under the conditions that at time $t=0$ the particles started at
$r=0$, with $\varphi$ uniformly distributed.  Integrating over the
angle $\varphi$ leads to the marginal distribution of displacements:
\begin{equation}
\rho(r)=\frac{r}{2 D_\text{eff}t}e^{-\frac{r^2}{4D_\text{eff}t}}.
\label{eq:rayl}
\end{equation}

\begin{figure}[ht]
  \includegraphics[width=.7\textwidth]{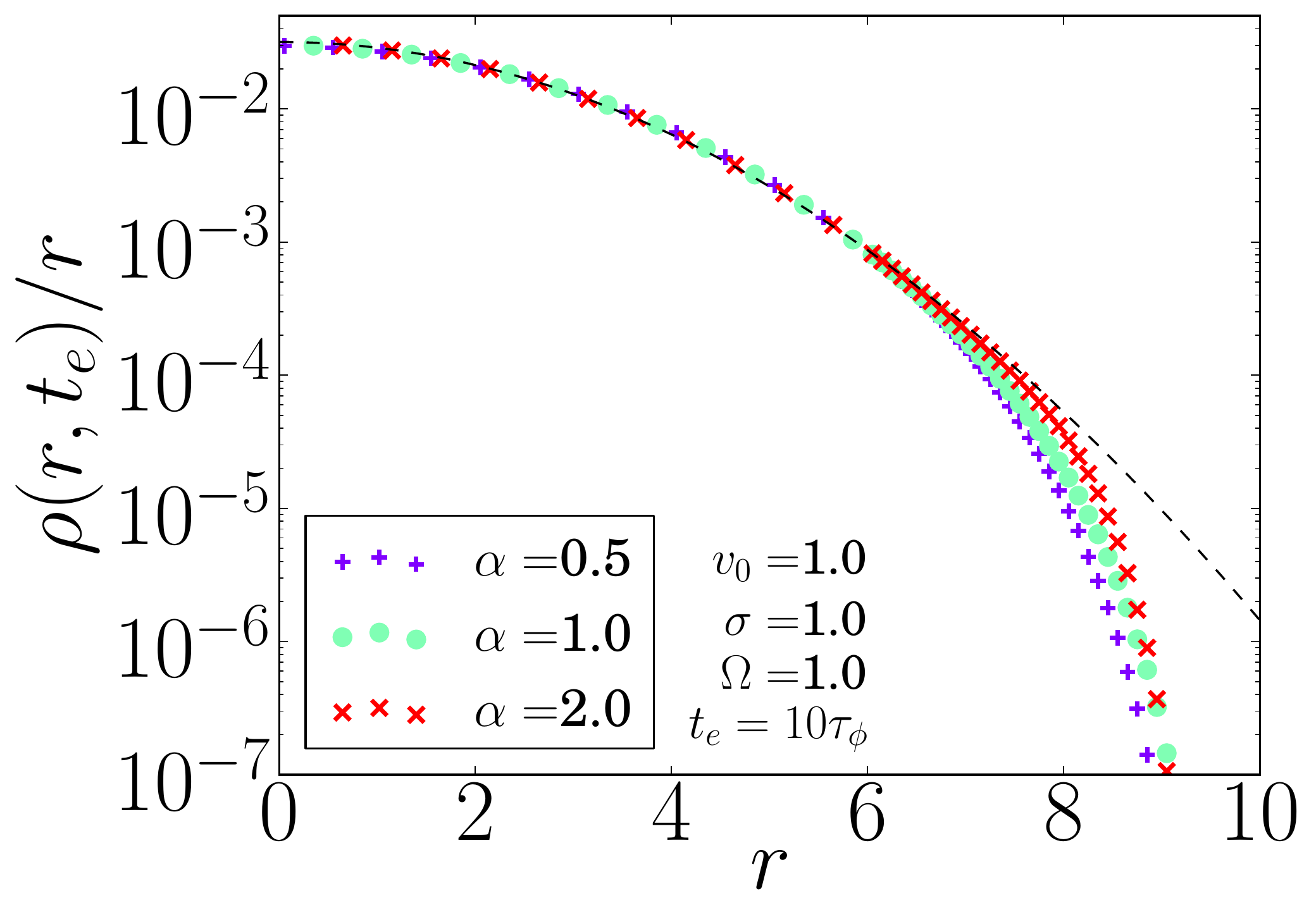}
  \caption{Colored symbols correspond to simulation results according
    to our full system \eqref{r_dot},\eqref{phi_dot} for the
    displacement $\rho(r,t_e)/r$ in polar coordinates divided by $r$,
    for different values of $\alpha$.  Black dashed line according to
    equation \eqref{eq:rayl} is the diffusion approximation.
    Parameters as given in the figures and $\tau_\phi=1$ for all curves.}
   \label{fig:disp}
\end{figure}
Figure \ref{fig:disp} shows the displacements' density for our initial
system and the results of our approximation. Symbols correspond to
simulation results for the initial system
Eqs.~\eqref{r_dot},\eqref{phi_dot}.  The black dashed line shows the
the Rayleigh distribution divided by $r$, for $t=10\tau_\phi$. For
$r\ll v_0t$ the approximation works well.  The curve starts to deviate
at distances larger then the mean squared displacement, which in this
case occurs when $r\, \approx \,6$.

Close to the maximal distance $r=v_0t$ the approximation breaks down
since it neglects the existence of the maximal absolute velocity and
therefore the truncation of the displacements'
distribution. \notizz{As can be seen from the simulation results, the
  index $\alpha$ of the noise influences weakly the exact form of the
  decay of $\rho(r,t)$ to
  zero.}

Thus, the coarse grained dynamics of our micro-swimmer with angular
component driven by symmetric $\alpha-$stable noise and constant
torque can be described by the standard Langevin equations
\begin{equation}
\frac{d\vec{r}}{dt}=\sqrt{2D_\text{eff}}\vec{\xi},
\label{eq:brown}
\end{equation}
where two independent Gaussian noise sources $\vec{\xi}=(\xi_x,\xi_y)$
drive the motion, with $\langle \xi_i \rangle=0$ and $\langle
\xi_i(t)\xi_j(t')\rangle=\delta_{ij}\delta(t-t')$, where
$i,j =x,y$.

v) In the limit $t \gg \tau_{\phi}$ the trajectories of
stochastic microswimmers with constant speed are indistinguishable
from the paths created by a Brownian particle as predicted by
Eq.~\eqref{eq:brown} with noise possessing $\alpha=2$. The question
what kind of noise drives the heading is then addressed to the
dependence of the diffusion coefficient on the parameter $\alpha$.
Following Eq.~\eqref{eq:deff} and Eq.~\eqref{eq:relaxt}, the diffusion
coefficient scales in case of strong noise as
\begin{equation}
D_\text{eff}\,\propto\, v_0^2 \, \tau_{\phi}\,=\,\frac{v_0^{2+\alpha}}{\sigma^{\alpha}} \,.
\label{eq:deff2}
\end{equation} 
The experimentally accessible value is the velocity $v_0$ of the
microswimmer, rather than the noise intensity $\sigma$ of the torque.
Hence, the inspection of the diffusion coefficient's dependence on the
velocity may give hints onto the possible presence of the
$\alpha$-stable noise. Furthermore, we emphasize that the diffusion
coefficient scales counterintuitively as $D_\text{eff}\,\propto
1/\sigma^{\alpha}$ with the noise intensity, parallel to what is known
for $\alpha=2$ \cite{mikhailov}.

vi) The influence of the torque induces an anti-symmetric part in the
matrix of the Onsager coefficients connecting the momentum flux with
the derivative of the density in Eq.~\eqref{eq:fluxes}.  The latter
results in a rotation matrix with the rotation angle $\phi_{\Omega}$
which is defined as $\sin(\phi_{\Omega} )= \Omega \tau_{\phi}/
\sqrt{1+\Omega^2 \tau_{\phi}^2}$.  In compact form the connections
Eq.~\eqref{eq:fluxes} read
\begin{eqnarray} 
\rho \vec{u}\,=\,- \frac{\tau_{\phi}}{\sqrt{1+\Omega^2 \tau_{\phi}^2}} \frac{v_0^2}{2}
\begin{pmatrix}
         ~\cos(\phi_{\Omega})~~~~-\sin(\phi_{\Omega}) \\ \sin(\phi_{\Omega})~~~~\cos(\phi_{\Omega}) \end{pmatrix}\,
\vec{\nabla} \rho\,.
\label{rhou_dreh}
\end{eqnarray}

In conclusion, we considered an active particle moving at a constant
speed, with constant torque and random fluctuations in the heading.
Such particles exhibit ballistic motion at small time scales
$t<\tau_\phi$ and diffusive behavior at larger times. We showed that
starting from the Fokker Planck equation for the joint probability
density of the position and the heading, the orientation as a
persistent variable can be eliminated by means of reduced moments for
all symmetric $\alpha$-stable noise sources.  This leads to the
diffusion equation for the corresponding coarse grained dynamics. In
consequence, the resulting particle dynamics becomes that of a
Brownian particle.

\section{Acknowledgments}
This work was supported by the Deutsche Forschungsgemeinschaft via
IRTG 1740. LSG thanks Ohio University in Athens OH and especially
A. B. Neiman for hospitality and support. The authors thank Vander
Freitas (Sao Jose dos Campos) for pointing out the behavior of fruit
flies. The authors thank Christophe Haynes for fruitful discussions.

\appendix
    \section{Adiabatic elimination using the Langevin equation for Gaussian white noise}\label{append1}
In the case of Gaussian white noise ($\alpha=2$), it is possible to
use the Langevin equations Eq.~\eqref{r_dot} and Eq.~\eqref{phi_dot}
to eliminate the persistent variable. We outline here this
approach for the sake of completeness. To be consistent with our
previous definition, we require
$\langle\xi(t)\xi(t')\rangle=2\delta(t-t')$.

Taking the time derivative of the velocity components 
\begin{equation}
  \dot{v}_x=\frac{v_0(\cos(\phi(t)+{\rm d}\phi)-\cos(\phi(t))}{{\rm d} t}=\frac{v_0(\cos(\phi)\cos({\rm d}\phi)-\sin(\phi)\sin({\rm d}\phi)-\cos(\phi))}{{\rm d} t}\,,
\end{equation}
(with  ${\rm d} t\rightarrow 0$) allows for a compact description. Stratonovich calculus \cite{stratonovich} will be used later on.  

According to Eq.~\eqref{phi_dot} the angular increment reads:
\begin{equation}
  {\rm d}\phi=\frac{1}{\tau_\phi}{\rm d} W_t+\Omega{\rm d} t, 
  \label{eq:ang_inc}
\end{equation}
with ${\rm d} W_t$ being the Wiener process, with the average
properties: $\langle{\rm d} W_t\rangle=0 $, $\langle{\rm d} W_t{\rm d}
W_{t'}\rangle=0 $ for $t\ne t'$ and $\langle{\rm d} W_t{\rm d}
W_{t'}\rangle=2{\rm d} t $ for $t=t'$. The non-averaged squared
increment behaves as $( {\rm d} W_t)^2\,=\,2{\rm d} t\,+\,
\mathcal{O}({\rm d} t^{3/2})$.  Taking ${\rm d}\phi$ small, the
change in velocity can be rewritten as
\begin{equation}    
  \dot{v}_x\,=\,-\frac{v_0(\cos(\phi)\frac{{\rm d} \phi^2}{2}+\sin(\phi){\rm d} \, \phi)}{{\rm d} t}\,.
\end{equation}
Up to first order of ${\rm d}t$ this equation reduces to
\begin{equation}    
\dot{v}_x=-\frac{1}{\tau_\phi}\left(v_x+\Omega \tau_\phi v_y+\sqrt{v_0^2\tau_\phi}\sin(\phi)\, \xi_\phi(t)\right).
\end{equation}
In this expression we extracted the relaxation time introduced in
Eq.~\eqref{eq:relaxt} for $\alpha=2$.

Now we can adiabatically eliminate the change in velocity. For times
larger than the relaxation time $t\gg\tau_\phi$, we reach the limit,
corresponding to $\dot{v}_x\tau_\phi \to 0$.  The velocity reads now:

\begin{equation}    
v_x\,=\,-\Omega \tau_\phi v_y-\sqrt{v_0^2\tau_\phi}\sin(\phi)\,\xi_\phi(t)\,.
\end{equation}
The velocity in the $y$ direction can be determined following the same lines:
\begin{equation}    
v_y\,=\,-\Omega \tau_\phi v_x-\sqrt{v_0^2\tau_\phi}\cos(\phi)\,\xi_\phi(t)\,.
\end{equation}
Both velocities depend on each other. Eliminating these dependencies,
i.e. substituting one equation into the other one, results in the
closed equations
\begin{equation}
  v_x\, =\, - \frac{1}{1+\left( \Omega \tau_{\phi}\right)^2}\, \left(\sin(\phi) \,+\,\Omega \tau_{\phi} \cos(\phi)\right)\,\sqrt{v_0^2\tau_\phi} \xi_{\phi}(t)
  \label{eq:velo_x}
\end{equation}
and
\begin{equation}
  v_y\, =\, \frac{1}{1+\left( \Omega \tau_{\phi}\right)^2}\, \left(\cos(\phi) \,-\,\Omega \tau_{\phi} \sin(\phi)\right)\,\sqrt{v_0^2\tau_\phi} \xi_{\phi}(t)\,.
  \label{eq:velo_y}
\end{equation}
We point out that the same noise $\xi_{\phi}(t)$ is acting in both projections of the velocity.  Further more, it should be noted that the angle $\phi(t)$ and the noise
$\xi_\phi(t)$ are statistically independent since the former depends only on the values of $\xi(t')$ at previous instants of time. 

We now can calculate the velocity correlation function $C_{v,v}(t-t^\prime)=\langle \vec{v}(t)\cdot \vec{v}(t^\prime)\rangle$ and obtain
\begin{equation}  
C_{v,v}(t-t^\prime)\,=\, 2\frac{v_0^2 \tau_\phi}{1+ \left(\Omega \tau_\phi\right)^2}\, \delta(t-t^\prime)\,.
\end{equation}
Thus, the velocity is given by a white noise and inertia has being
eliminated.

Nevertheless, the velocity components are still correlated since a
single noise source acts in both directions. Only after averaging over
the angle $\phi$ with the uniform angular distribution following from
equation Eq.~\eqref{eq:pphi}, for $t\gg \tau_{\phi}$, do the components
become uncorrelated;
\begin{equation}   
\langle\langle v_i(t)v_j(t')
\rangle\rangle_\phi\,=\,\frac{v_0^2\tau_\phi}{1+\left(\Omega \tau_\phi \right)^2}\,\delta(t-t')\,\delta_{ij}\,.
\end{equation}
In this limit the dynamics can be formulated as defined in
Eq.~\eqref{eq:brown}, and corresponds to those of a Brownian particle
driven by two independent noise sources in both directions.  The
correlated dynamics Eq.~\eqref{eq:velo_x} and Eq.~\eqref{eq:velo_y}
still reflects the fact that the noise $\xi_\phi(t)$ is acting in a
direction perpendicular to the current motion.  In case of an
additionally acting torque $\Omega$ the two forces undergo an
additional shift which is given by the angle $\phi_\Omega$ as
defined in the last section.  Using this and Eq.~\eqref{eq:velo_x} and
Eq.~\eqref{eq:velo_y} the two components reads
\begin{eqnarray}
&&\dot{x}\,=\,v_x\,=\,-\sqrt{•2 D_{\rm eff}} \, \sin(\phi(t)+\phi_\Omega) \,\xi_\phi(t) \nonumber \\
&&\dot{y}\,=\,v_y\,=\,\sqrt{•2 D_{\rm eff}} \, \cos(\phi(t)+\phi_\Omega) \,\xi_\phi(t)\,.
\label{eq:overdamped}
\end{eqnarray}  
We remember that $\phi(t)$ defines the heading, i.e. the current
direction of motion. In Eq.~\eqref{eq:overdamped} the velocity or the
displacement points perpendicular to the direction given by the angle
$\phi(t)+\phi_\Omega$. In particular, without torque $\Omega=0$, it
acts perpendicular to the current motion.

Eq.~\eqref{eq:overdamped} can be interpreted as an algorithm for the
considered coarse grained microswimmer. For time scales
$t\gg\tau_\phi$ both $\phi(t)$ and $\xi_\phi(t)$ are statistically
independent. $ \phi(t)$ is a white noise homogeneously distributed in
$[0,2\pi]$ and $\xi_\phi(t)$ is Gaussian white noise.  The process
$\phi(t)$ after the elimination procedure does not possess any memory
and is in the considered case of $\alpha=2$ the increment of the
Wiener process. Therefore in the approximation, the microswimmer
changes suddenly the direction of motion following a Wiener process
shifted by $\phi_\Omega$ and increments are added in direction
perpendicular to $\phi(t)+\phi_\Omega$.

\notizz{
\section{Moments of the angular dynamics: }
\label{frac_mom}
In contrast to the consideration in the main part of this article, we
start here with the pdf $\tilde{P}(x,
\phi,t|x_0,\phi_0,t_0)$, defined for $\phi \in
(-\infty,+\infty)$. Later on, we again omit the explicit statement of the condition. 

The connection between the the unwrapped and the wrapped distribution
function $P(x,\phi,t)$  can be
formulated as follows
\begin{equation}
P(x,\phi,t)\,=\,\sum_{n=-\infty}^{\infty}\, \tilde{P}(x, \phi + 2\pi
n,t)\,,
\label{eq:wrapping}
\end{equation}
where now $\phi \in [0, 2\pi)$ and the wrapped pdf is normalized in
this interval.
  
As $\tilde{P}(x,\phi,t)$ is a probability density, it
fulfills
\begin{equation}
\tilde{P}(x,\phi,t)\ge0, 
\end{equation}
and
\begin{equation}
  \int_{-\infty}^{\infty}\, {\rm d}\phi\, \tilde{P}(x,\phi,t) \,<\,\infty;\, \,\,\forall x,t.
\end{equation}
The expectation values $ \langle \cos^n(\phi)\rangle, ~~n \,\ge\, 1$ will
exist as well since the following holds
\begin{equation}
  \left|\int_{-\infty}^{\infty}\, {\rm d}\phi\, \cos^n(\phi)
  \,\tilde{P}(x,\phi,t) \right|\, \le\, \int_{-\infty}^{\infty} \,
          {\rm d}\phi \, \tilde{P}(x,\phi,t)\,,
\end{equation}
as $|\cos^n(\phi)|\le 1$ is bounded. Similarly, we proceed with $\langle
\sin(\phi)^n\rangle$.

During the derivation of the transport equations we have used the
following identities and properties of the pdf
\begin{eqnarray}
&& \int_{-\infty}^{\infty} \, {\rm d}\phi \,
  \cos(\phi)\,\frac{\partial^\alpha}{\partial|\phi|^\alpha}\,
  \tilde{P}(x,\phi,t)\,
  =\,\frac{1}{2\pi}\,\int_{-\infty}^{\infty}\int_{-\infty}^{\infty}\,
  {\rm d}k {\rm d}\phi
  \,\cos(\phi)\,\frac{\partial^\alpha}{\partial|\phi|^\alpha}e^{-ik\phi}\,\tilde{P}(x,k,t)
  \nonumber \\ &&
  =\,-\,\frac{1}{2\pi}\int_{-\infty}^{\infty}\int_{-\infty}^{\infty}\,
  {\rm d}k {\rm
    d}\phi|\,k|^\alpha\cos(\phi)e^{-ik\phi}\,\tilde{P}(x,k,t)
  \,=\,-\,\frac{1}{2}\left(\tilde{P}(x,1,t)\,+\,\tilde{P}(x,-1,t)\right)\,.\nonumber
\end{eqnarray}
where $\tilde{P}(x,k,t)$ is the Fourier transform of $\tilde{P}(x,\phi,t)$ in its
angular variable. Otherwise, it holds that 
\begin{eqnarray}
\int_{-\infty}^{\infty}{\rm d}\phi \,
  \cos(\phi)\,\tilde{P}(x,\phi,t) &=&\frac{1}{2\pi}\,
  \int_{-\infty}^{\infty}\int_{-\infty}^{\infty}\,{\rm d}k {\rm d}\phi
  \, \cos(\phi)e^{-ik\phi} \,\tilde{P}(x,k,t)\nonumber \\ &=&\frac{1}{2}\,\left(\tilde{P}(x,1,t)\,+\,\tilde{P}(x,-1,t)\right)\,.\nonumber
\end{eqnarray}
It follows that
\begin{equation}
  \int_{-\infty}^{\infty}\,{\rm d}\phi
  \,\cos(\phi)\,\frac{\partial^\alpha}{\partial|\phi|^\alpha}\,\tilde{P}(x,\phi,t)
  \,=\,-\,\int_{-\infty}^{\infty} {\rm d}\phi \,
  \cos(\phi)\,\tilde{P}(x,\phi,t)\,.
\end{equation}

Analogously, we derive the identities:
\begin{equation}
  \int_{-\infty}^{\infty} {\rm d}\phi\, \sin(\phi)\frac{\partial^\alpha}{\partial|\phi|^\alpha}\tilde{P}(x,\phi,t) \,=\,-\int {\rm d}\phi\, \sin(\phi)\tilde{P}(x,\phi,t)\,.
\end{equation}

\begin{eqnarray}
&& \int_{-\infty}^{\infty} {\rm d}\phi\,
  \cos^2(\phi)\,\frac{\partial^\alpha}{\partial|\phi|^\alpha}\,\tilde{P}(x,\phi,t)
  \,=\,-2^\alpha\,\int_{-\infty}^{\infty} {\rm
    d}\phi\,\left(\cos^2(\phi)\,-\,\frac{1}{2}\,\right)\,\tilde{P}(x,\phi,t)\,,\nonumber
  \\ &&\int_{-\infty}^{\infty} {\rm d}\phi\,
  \sin^2(\phi)\,\frac{\partial^\alpha}{\partial\,|\phi|^\alpha}\,\tilde{P}(x,\phi,t)
  \,=\,2^\alpha\int_{-\infty}^{\infty} {\rm d}\phi\,
  \left(\cos^2(\phi)\,-\,\frac{1}{2}\right)\,\tilde{P}(x,\phi,t)\,,\nonumber
  \\ && \int_{-\infty}^{\infty} {\rm d}\phi\,
  \cos(\phi)\sin(\phi)\,\frac{\partial^\alpha}{\partial|\phi|^\alpha}\,\tilde{P}(x,\phi,t)
  \,=\,-2^\alpha\,\int_{-\infty}^{\infty} {\rm
    d}\phi\,\cos(\phi)\sin(\phi)\,\tilde{P}(x,\phi,t) \,.\nonumber
\end{eqnarray}
The derived relations express properties of the generating function of
the $\alpha$-stable noise. They hold as well for the wrapped
distribution with different integration limits due to the linear
connection \eqref{eq:wrapping} between both presentations.
}

\end{document}